%% file: tensorpol_main.tex
\documentclass[preprint,showpacs,preprintnumbers,amsmath,amssymb]{revtex4-1}

\usepackage{graphicx}
\usepackage{dcolumn}
\usepackage{bm}

\begin{document}

\title{Hyperfine-mediated static polarizabilities of monovalent atoms and ions}

\newcommand{\Reno}{
Department of Physics, University of Nevada, 
Reno, Nevada 89557, USA}
\newcommand{\Sydney}{
School of Physics, University of New South Wales,
Sydney, NSW 2052, Australia}
\newcommand{\Auckland}{
Centre for Theoretical Chemistry and Physics, 
The New Zealand Institute for Advanced Study, 
Massey University Auckland, Private Bag 102904, 0745, Auckland, New Zealand}

\author{V. A. Dzuba}
\affiliation{\Sydney}

\author{V. V. Flambaum}
\affiliation{\Sydney}

\author{K. Beloy}
\affiliation{\Auckland}

\author{A. Derevianko}
\affiliation{\Reno}

\begin{abstract}
We apply relativistic many-body methods to compute static
differential polarizabilities for transitions inside the ground-state hyperfine manifolds of monovalent atoms and ions. Knowing
this transition polarizability is required in a number of high-precision experiments, such as microwave
atomic clocks and searches for CP-violating permanent electric dipole moments.
While the traditional polarizability arises in the second-order of interaction with the externally-applied
electric field,  the differential polarizability involves  additional contribution from the hyperfine interaction of atomic electrons with nuclear moments.
We derive formulas for the scalar and tensor polarizabilities including contributions from magnetic dipole and electric quadrupole hyperfine interactions.  Numerical results are presented for Al, Rb, Cs, Yb$^+$, Hg$^+$, and Fr.
\end{abstract}

\pacs{31.15.Ar,31.25.-v,32.60.+i}
\maketitle

\date{\today}

\input{intro} 
\input{theory} 
\input{numerics_mbpt}
\input{results_discussion} 

\section*{Acknowledgments}
The work of A.D. was supported in part by the U.S. NSF and by U.S. NASA under Grant/Cooperative
Agreement No. NNX07AT65A issued by the Nevada NASA EPSCoR program.
The work of V.A.D. and V.V.F was supported in part by the ARC. The work of K.B. was supported by the Marsden Fund of the Royal Society of New Zealand.


%

\end{document}

%% file: intro.tex
\section{Introduction}
When an atom is placed in an external electric field, its energy levels shift due to the Stark effect. For states of definite parity, the effect arises in the second order in the interaction of atomic electrons with the external E-field. The energy shift $\delta E_a$ is conventionally parameterized in terms of the polarizability of the atomic state $\alpha_a$,
\begin{equation}
\delta E_a = -\frac{1}{2} \alpha_a \mathcal{E}_0^2 \, ,
\label{eq:alpha-E}
\end{equation}
where $\mathcal{E}_0$ is the strength of the applied E-field.

The polarizability depends on atomic electric-dipole $D$ matrix elements
and energies $E$
\begin{equation}
\alpha_a=
2\,  \sum_{b\neq a}\frac{
\left\vert \langle a|D_z |b\rangle\right\vert ^{2}}{E_{b}- E_{a}} \, .
\label{Eq:PolarizabilityGeneral}%
\end{equation}
The sums are over a complete atomic eigen-set and the $z$-axis has been chosen along the E-field. On general grounds,
we may decompose the polarizability for a state $|n FM_F\rangle$ of the total angular momentum $F$ and its projection $M_F$ into the following contributions,
\begin{equation}
\alpha_{nFM_F}=  \alpha_{nF}^{S}
+       \frac{3{M_{F}}^{2}-F(F+1)}{F(2F-1)}\alpha_{nF}^{T}   \, .
\label{Eq:alphaBreakDown}
\end{equation}
Here the superscripts $S$ and $T$ distinguish the scalar and tensor parts of the polarizability. The ``polarizabilities'' $\alpha_{nF}^{S}$ and $\alpha_{nF}^{T}$ no longer depend on the magnetic quantum number $M_F$.

In this paper we focus on a difference of polarizabilities between two states $nF'$ and $nF$ of the same hyperfine manifold of states of total orbital angular momentum $J=1/2$. Such calculations require additional care. Indeed, we are considering the Stark shift of hyperfine levels attached to the same electronic
state. To the leading order, the shift is determined by the properties of the underlying electronic state. However, because the electronic state for
both hyperfine levels is the same, the scalar Stark shift of both levels is the same.
An apparent difference between the two levels is caused by the hyperfine interaction (HFI), and the rigorous analysis involves so-called HFI-mediated polarizabilities~(see, e.g., \cite{RosGheDzu09}). Similar arguments hold for the tensor part of the polarizability. $\alpha_{nF}^{T}$, taken with its prefactor in Eq.~(\ref{Eq:alphaBreakDown}
), is an expectation value of an irreducible tensor operator of rank 2; it simply vanishes for $J=1/2$ states due to the angular selection rules. Only the HFI coupling of nuclear and electronic momenta ($\mathbf{F}=\mathbf{I}+\mathbf{J}$) leads to nonzero values of the tensor polarizability.

Early works on Stark shifts of transition frequencies within hyperfine manifolds include Refs.~\cite{HauZac57,LipSan64,San67,CarAdlBak68,GouLipWei69}.
More recent interest to this problem was motivated by the Stark shifts of the hyperfine transition frequency due to the ambient black-body radiation (BBR)~\cite{ItaLewWin82}. The BBR shift is a major systematic correction in microwave clocks, especially the $^{133}$Cs primary frequency standard~\cite{WynWey05}. This motivated the most precise measurement of the DC Stark shift in a Cs fountain~\cite{SimLauCla98}. The relevant Stark shifts were a subject of many recent works (see, e.g., state-of-the-art calculations Ref.~\cite{BelSafDer06,AngDzuFla06PRL} and references therein).

Perhaps the most complete earlier theoretical treatment within the third-order (two electric-dipole couplings and one HFI) perturbation theory was given by Sandars \cite{San67} in 1967. However, only recently (i.e., four decades later), a sign mistake in the expression of Ref.~\cite{San67}
for the tensor part of the HFI-mediated polarizability was discovered~\cite{UlzHofMor06,HofMorUlz08} (a correct result for tensor polarizability of Tl was obtained earlier in Ref.~\cite{SkoFla78}).

This sign error is directly relevant to extracting BBR correction from high-precision experiments.
Notice that due to the isotropic nature of the BBR, the BBR clock shift is expressed in terms of the {\em scalar} part of the HFI-mediated polarizability. Moreover, characteristic frequencies of room-temperature BBR are well below excitation energies of atomic transitions thereby justifying replacing frequency-dependent polarizability with  DC polarizability~\cite{ItaLewWin82}.
Accordingly, the modern value of the BBR correction for the Cs clock is based on
a measurement~\cite{SimLauCla98} which was carried out in a DC E-field. However, the measured  Stark shift involves a combination~(\ref{Eq:alphaBreakDown}) of both scalar and tensor polarizabilities. Therefore to arrive at the BBR shift, one needs to remove the contribution due to the tensor polarizability. Clearly, the sign mistake discovered in~\cite{UlzHofMor06,HofMorUlz08} becomes relevant.

Here we  extend our earlier treatment of the HFI-mediated polarizabilities~\cite{BelSafDer06,AngDzuFla06PRL,RosGheDzu09} with a specific focus on the tensor polarizabilities. Compared to Refs.~\cite{UlzHofMor06,HofMorUlz08} we employ a fully-relativistic formalism  and evaluate tensor polarizabilities for several atoms and ions using modern relativistic many-body methods. We independently confirm that indeed, Ref.~\cite{San67} had a sign mistake, requiring reinterpretation of measurements~\cite{SimLauCla98}. We also evaluate the tensor polarizability for the secondary frequency standard based on Rb atoms.

Another motivation for our work comes from searches for the so far elusive permanent electric-dipole moments (EDM). Non-vanishing EDMs violate both time- and parity-reversal symmetries.  Planned experiments will be carried out with Fr atoms~\cite{FrEDM}. These atoms will be placed in a strong electric field and so-far unknown $M_F$-dependent tensor polarizabilities would contribute to the error budget of the EDM search. Our computed values for Fr isotopes will aid the design and interpretation of these planned experiments.

The paper is organized as follows. In Section~\ref{Sec:Theory} we derive fully-relativistic third-order formulae for HFI-mediated tensor polarizabilities. Section~\ref{Sec:Calculations} presents details of numerical evaluation within relativistic many-body theory. Finally, the results are discussed and compared with literature values in Section~\ref{Sec:Results}. Unless specified otherwise, atomic units, $m_e=\hbar=|e|=1$ are used throughout. 

%% file: theory.tex
\section{Theoretical setup}
\label{Sec:Theory}

We are interested in transitions between two hyperfine components of the same
electronic states. Below we employ the conventional labeling scheme for the
atomic eigenstates, $\left\vert n\left( IJ\right)  FM_{F}\right\rangle $,
where $I$ is the nuclear spin, $J$ is the electronic angular momentum, and $F$
is the total angular momentum, $\mathbf{F}=\mathbf{J}+\mathbf{I}$. $M_{F}$ is
the projection of $F$ on the quantization axis and $n$ encompasses the
remaining quantum numbers.

\begin{figure}[h]
\begin{center}
\includegraphics*[scale=0.35]{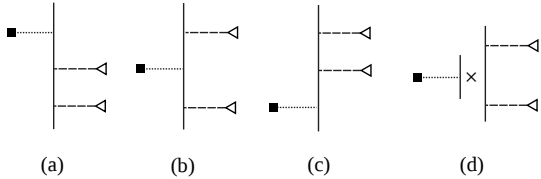}
\caption{Diagrams representing the complete third order hyperfine-mediated polarizability. Each diagram contains a hyperfine interaction (dotted line capped with a filled square) in addition to two interactions with the external electric field (dashed line capped with an empty triangle). The diagrams correspond to the (a) top, (b) center, (c) bottom, and (d) normalization terms discussed in the text.   
\label{Fig:diagrams}}
\end{center}
\end{figure}

As discussed in the introduction, computation of the transition polarizability for $J=1/2$ hyperfine-manifolds requires third-order analysis. This involves two perturbations due to the externally-applied electric field, $V_E = -\mathbf{D} \cdot \mathcal{E}$, and hyperfine 
interaction $V_\mathrm{HFI}$. These perturbations may be chained into three distinct 
diagrams (see Fig.~\ref{Fig:diagrams}). Additionally, there is a residual (or normalization) diagram.
  
We would like to stress the importance of a consistent treatment of the  HFI-mediated polarizabilities (i.e., including all the diagrams in Fig.~\ref{Fig:diagrams}).  Consider a general expression for the scalar polarizability,
\begin{equation}
\alpha_{nF}^S = \frac{2}{3}\sum_{k=x,y,z}\sum_{i=|n_{i}F_{i}M_{i}\rangle }\frac{\langle nFM_{F}|~D_{k}|i\rangle \langle i~|D_{k}|nFM_{F}\rangle }{E_{nFM_{F}}-E_{i}}.
\end{equation}
Here all the involved states are the hyperfine
states. While this requires that  the energies include hyperfine splittings, it also  means
that the wave-functions incorporate HFI to all orders of perturbation theory.
Including the experimentally-known hyperfine splittings in the
summations is straightforward but limiting ourselves to this approximation would exclude the HFI corrections to the wave-functions.
By expanding the energy denominators, we observe that including HFI into energies would only recover the residual diagram and partially the center diagram.
We find that the remaining contributions are of the same order, and limiting computations to HFI-induced energy shifts only is hardly justified.

Previously, we derived equations for {\em dynamic} HFI-mediated polarizabilities of hyperfine states 
in Ref.~\cite{RosGheDzu09}. Clearly, {\em static} polarizabilities can be obtained by setting laser frequency to zero in the derived formulas. There is, however, one important addition to the formulae presented in Ref.~\cite{RosGheDzu09}: the HFI operator in that paper was truncated at the magnetic-dipole interaction. Here we additionally include the HFI coupling due to the electric-quadrupole nuclear moment. This contribution to tensor polarizabilities becomes increasingly important for heavier atoms.
Note that the electric quadrupole contribution to scalar polarizabilities
and thermal shift
of states with total momentum $j=1/2$ is zero. This can be explained in 
the following way.
Scalar polarizability can be separated from total energy shift by averaging 
over directions of electric field.
One cannot make non-zero scalar (energy shift is a scalar)
from remaining electron angular momentum 1/2
and nuclear quadrupole, $j_a j_b Q_{ab}=\sigma_a \sigma_b Q_{ab}=0$
(squared Pauli matrix sigma is reduced to delta symbol and  
antisymmetric linear tensor with $\sigma$,
$Q_{ab}$  is symmetric with zero trace, so that $\sum_a Q_{aa}=0$).

On general grounds, the rotationally-invariant hyperfine interaction between atomic electrons 
and nuclear moments may be written as a sum over scalar products of irreducible tensor operators (we follow notation of Ref.~\cite{Joh07book})
\begin{equation}
V_\mathrm{HFI}=\sum_{N}\mathcal{N}^{\left(  N\right)  }\cdot \mathcal{T}^{\left(  N\right) } \,.
\end{equation}
Here the  irreducible tensor operators  $\mathcal{N}^{\left(  N\right)  }$ and $\mathcal{T}^{\left(
N\right)  }$ act in the space of nuclear and electronic coordinates
respectively, with $N$ being their ranks. The nuclear magnetic moment is
conventionally defined as
\begin{equation}
\mu=\langle I,M_{I}=I|\mathcal{N}_{0}^{\left(  1\right)  }|I,M_{I}=I\rangle
\end{equation}
and the nuclear electric-quadrupole moment as 
\begin{equation}
Q=2\langle I,M_{I}=I|\mathcal{N}_{0}^{\left(  2\right)  }|I,M_{I}=I\rangle \, .
\end{equation}
In the formulas below we
require the reduced matrix element of the nuclear moment operator in the
nuclear basis. For magnetic dipole, this is related to the
nuclear magnetic $g$-factor as
\[
\langle I||\mathcal{N}^{\left(  1\right)}||I\rangle=\frac{1}{2}\sqrt{\left(  2I\right)  \left(
2I+1\right)  \left(  2I+2\right)  }~g\mu_{n},
\]
$\mu_{n}$ being the nuclear magneton and $\mu=gI\mu_n$. For the electric-quadrupole moment,
\[
\langle I||\mathcal{N}^{\left(  2\right)  }||I\rangle=\frac{\sqrt{\left(
2I-2\right)  !\left(  2I+3\right)  !}}{2\left(  2I\right)  !}\, Q \, .
\]

The components of relevant electronic tensors are
\begin{align*}
\mathcal{T}_{\lambda}^{\left(  1\right)  } &  =-\frac{i\sqrt{2}\left(  \mathbf{\alpha}\cdot
\mathbf{C}_{1\lambda}^{\left(  0\right)  }\left(  \mathbf{\hat{r}}\right)
\right)  }{cr^{2}} \, , \\
\mathcal{T}_{\lambda}^{\left(  2\right)  } &  =-\frac{C_{\lambda}^{2}\left(  \mathbf{\hat{r}}\right)
}{r^{3}} \, ,
\end{align*}
where $r$ is the electronic coordinate, $\mathbf{\alpha}$ are the Dirac matrices, and $\mathbf{C}_{1\lambda}^{\left(  0\right)  }\left(  \mathbf{\hat{r}}\right)$ and $C_{\lambda}^{2}\left(  \mathbf{\hat{r}}\right)$ are normalized vector spherical harmonic and normalized spherical harmonic functions, respectively~\cite{VarMosKhe88}.

A derivation similar to Ref.~\cite{RosGheDzu09} results in
the scalar and  the tensor polarizabilities given by (here $[F]=2F+1$)%
\begin{align*}
\alpha_{nF}^{s}  &  =\frac{1}{\sqrt{3}}\frac{1}%
{\sqrt{[F]}}\alpha_{nF}^{\left(  0\right)  } ,\\
\alpha_{nF}^{T}  &  =-\left[  \frac{2}{3}\frac{~\left(
2F\right)  \left(  2F-1\right)  }{\left(  2F+1\right)  \left(  2F+2\right)
\left(  2F+3\right)  }\right]  ^{1/2}\alpha_{nF}^{\left(  2\right)  } \, ,
\end{align*}
with  the reduced polarizabilities $\alpha_{nF}^{\left(  K\right)}$ being sums over values of individual
diagrams of Fig.~\ref{Fig:diagrams},
\begin{equation}
\alpha_{nF}^{\left(  K\right)  }  =\sum_{N} (2\alpha
_{nF;N}^{\left(  K\right)  }\left( \mathrm{T}\right)  +\alpha
_{nF;N}^{\left(  K\right)  }\left( \mathrm{C}\right)  +\alpha
_{nF;N}^{\left(  K\right)  }\left(\mathrm{O}\right)) \,, 
\end{equation}
where we used the equality of the top and bottom diagrams.

The angular reduction of individual diagrams leads to expressions
\begin{align*}
\alpha_{nF;N}^{\left(  K\right)  }\left(  \mathrm{T}\right)   &
=[F]\sqrt{[K]}\\
&  \sum_{J_{a}J_{b}}\left(  -1\right)  ^{J+J_{a}}\left\{
\begin{tabular}
[c]{lll}%
$I$ & $I$ & $N$\\
$J_{a}$ & $J$ & $F$%
\end{tabular}
\ \ \ \right\}  \left\{
\begin{tabular}
[c]{lll}%
$J$ & $1$ & $J_{b}$\\
$1$ & $J_{a}$ & $K$%
\end{tabular}
\ \ \ \right\}  \left\{
\begin{tabular}
[c]{lll}%
$K$ & $J$ & $J_{a}$\\
$I$ & $F$ & $F$%
\end{tabular}
\ \ \ \ \right\}  T_{J_{a}J_{b}}^{\left(  K\right)  }\left(  nJ,N\right),
\end{align*}
\begin{align*}
\alpha_{nF;N}^{\left(  K\right)  }\left(  \mathrm{C}\right)  =[F]\sqrt{[K]} &
\sum_{J_{a}J_{b}}\sum_{J_{i}}[J_{i}](-1)^{2J_{a}+J_{b}+J+N-1}\times\\
&  \left\{
\begin{array}
[c]{ccc}%
J & J & J_{i}\\
I & I & N\\
F & F & K
\end{array}
\right\}  \left\{
\begin{array}
[c]{ccc}%
J & J & J_{i}\\
J_{a} & J_{b} & N\\
1 & 1 & K
\end{array}
\right\}  C_{J_{a}J_{b}}^{\left(  K\right)  }\left(  nJ,N\right),
\end{align*}
\begin{align*}
\alpha_{nF;N}^{\left(  K\right)  }\left(  \mathrm{O}\right)   &  =\left(
-1\right)  ^{2J+1}\left\{
\begin{array}
[c]{ccc}%
N & I & I\\
F & J & J
\end{array}
\right\}  \langle nJ||\mathcal{T}^{\left(  N\right)  }||nJ\rangle\langle
I||\mathcal{N}^{\left(  N\right)  }||I\rangle\times\\
&  \left[  F\right]  \sqrt{\left[  K\right]  }\left\{
\begin{array}
[c]{ccc}%
J & J & K\\
F & F & I
\end{array}
\right\}  \sum_{J_{a}}\left\{
\begin{array}
[c]{ccc}%
K & J & J\\
J_{a} & 1 & 1
\end{array}
\right\}  \times O_{J_{a}}^{\left(  K\right)  }\left(  nJ\right).
\end{align*}

Finally, the universal (these are independent on $F$, i.e., the clock
level) reduced sums are 
\begin{equation}
T_{J_{a}J_{b}}^{\left(  K\right)  }\left(  nJ,N\right)  =2\langle
I||\mathcal{N}^{\left(  N\right)  }||I\rangle\sum_{n_{a},n_{b}\neq
n}\frac{\langle nJ||\mathcal{T}^{\left(  N\right)  }||n_{a}J_{a}\rangle\langle
n_{a}J_{a}||D||n_{b}J_{b}\rangle\langle n_{b}J_{b}||D||nJ\rangle}{\left(
E-E_{a}\right)  \left(  E-E_{b}\right)  } \, ,
\label{Eq:TopSum}
\end{equation}
\begin{equation}
C_{J_{a}J_{b}}^{\left(  K\right)  }\left(  nJ,N\right)  =2\langle
I||\mathcal{N}^{\left(  N\right)  }||I\rangle\sum_{n_{a},n_{b}\neq n}\frac{\langle
nJ||D||n_{a}J_{a}\rangle\langle n_{a}J_{a}||\mathcal{T}^{\left(  N\right)
}||n_{b}J_{b}\rangle\langle n_{b}J_{b}||D||nJ\rangle}{\left(  E-E_{a}\right)
\left(  E-E_{b}\right)  } \, ,
\label{Eq:CenterSum}
\end{equation}
\begin{equation}
O_{J_{a}}^{\left(  K\right)  }\left(  nJ\right)  =2\sum_{n_{a}\neq n}\frac{\langle
nJ||D||n_{a}J_{a}\rangle\langle n_{a}J_{a}||D||nJ\rangle}{\left(
E-E_{a}\right)  ^{2}}\, .
\label{Eq:NormSum}
\end{equation}
In these sums the values of the total orbital momenta of intermediate states $J_a$ and $J_b$
are fixed. 

%% file: numerics_mbpt.tex
\section{Numerical evaluation}

\label{Sec:Calculations}
To perform the calculations we use an {\em ab initio} approach
which has been described in detail in Ref.~\cite{AngDzuFla06}.
In this approach high accuracy is attained by including
important many-body and relativistic effects.

Calculations start from the relativistic Hartree-Fock (RHF) method
in the $V^{N-1}$ approximation. This means that the initial RHF
procedure is done for a closed-shell atomic core with the valence
electron removed. After that, the states of the external electron are
calculated in the field of the frozen core. Correlations are
included by means of the correlation potential method~\cite{DzuFlaSil87}.
We use the all-order correlation potential $\hat \Sigma$ for Rb, Cs,
and Fr and second-order correlation potential $\hat \Sigma^{(2)}$ for
Al, Yb$^+$, and Hg$^+$. The all-order $\hat \Sigma$ 
includes two classes of the higher-order
terms: screening of the Coulomb interaction and hole-particle
interaction (see, e.g.,~\cite{DzuFlaSus89} for details).

To calculate $\hat \Sigma$ and $\hat \Sigma^{(2)}$ we need a complete
set of single-electron 
orbitals. We use the B-spline technique~\cite{JohBluSap88} to
construct the basis. The orbitals are built as linear combinations of
40 B-splines of order 9 in a cavity of radius 40$a_B$.
The coefficients are chosen from the condition that the
orbitals are the eigenstates of the RHF Hamiltonian $\hat H_0$ of the
closed-shell core. The all-order $\hat \Sigma$ operator is
calculated with the technique which combines solving equations for
the Green functions (for the direct diagram) with the summation over
complete set of states (exchange diagram)~\cite{DzuFlaSus89}.
The second-order $\hat \Sigma^{(2)}$ operator is calculated using
direct summation over complete set of states.

The correlation potential $\hat \Sigma$ is then used to build a new
set of single-electron states, the so-called Brueckner orbitals.
This set is to be used in the summation in equations (\ref{Eq:TopSum}),
(\ref{Eq:CenterSum}) and (\ref{Eq:NormSum}). Here again we use the B-spline
technique to build the basis. The procedure is very similar to
the construction of the RHF B-spline basis. The only difference is that
new orbitals are now the eigenstates of the $\hat H_0 + \hat \Sigma$
Hamiltonian.



Matrix elements of the HFI and electric dipole operators are found by
means of the time-dependent Hartree-Fock (TDHF)
method~\cite{DzuFlaSil87,DzuFlaSus84}. This method is equivalent 
to the well-known random-phase approximation (RPA). In the TDHF
method, the single-electron wave functions are presented in the form
$\psi = \psi_0 + \delta \psi$, where $\psi_0$ 
is the unperturbed wave function. It is an eigenstate of the RHF
Hamiltonian $\hat H_0$: $(\hat H_0 -\epsilon_0)\psi_0 = 0$.  $\delta
\psi$ is the correction due to external field. It can be found be
solving the TDHF equation 
\begin{equation}
    (\hat H_0 -\epsilon_0)\delta \psi = -\delta\epsilon \psi_0 - \hat F \psi_0 -
  \delta \hat V^{N-1} \psi_0,
  \label{TDHF}
\end{equation}
where $\delta\epsilon$ is the correction to the energy due to external field
($\delta\epsilon\equiv 0$ for the electric dipole operator), $\hat F$ is the
operator of the external field
($V_\mathrm{HFI}$ or $-\mathbf{D}\cdot \mathcal{E}$), and
$\delta \hat V^{N-1}$ is the correction to the self-consistent
potential of the core due to external field. 

The TDHF equations are solved self-consistently for all states in the
core. Then the matrix elements between any (core or valence) states
$n$ and $m$ are given by 
\begin{equation}
    \langle \psi_n | \hat F + \delta \hat V^{N-1} | \psi_m \rangle.
    \label{mel}
\end{equation}

The best results are achieved when $\psi_n$ and $\psi_m$ are the Brueckner
orbitals computed with 
the correlation potential $\hat \Sigma$.

We use equation (\ref{mel}) for all HFI and electric dipole matrix
elements in evaluating the top, bottom, and center diagrams
(Eqs.~(\ref{Eq:TopSum}),(\ref{Eq:CenterSum}),(\ref{Eq:NormSum})) 
except for the ground state HFI matrix element in the normalization diagram
where we use experimental data. The results are presented in section
\ref{Sec:Results}.  

%% file: results_discussion.tex
\section{Results and Discussion}
\label{Sec:Results}
\begin{table*}
\caption{Third-order hyperfine static polarizabilities of single-valence atoms.}
\label{t:1}
\begin{ruledtabular}
\begin{tabular}{rlr cllc rrrrr}
\multicolumn{1}{c}{$Z$} &
\multicolumn{1}{c}{Atom} &
\multicolumn{1}{c}{$A$} &
\multicolumn{1}{c}{$I$} &
\multicolumn{1}{c}{$\mu/\mu_N$\footnotemark[1]} &
\multicolumn{1}{c}{$Q\footnotemark[1]$} &
\multicolumn{1}{c}{$F$} &
\multicolumn{1}{c}{$\alpha^{S}$} &
\multicolumn{1}{c}{$\alpha_A^{T}$\footnotemark[2]} &
\multicolumn{1}{c}{$\alpha_B^{T}$\footnotemark[3]} &
\multicolumn{1}{c}{$\alpha^{T}$\footnotemark[4]}  &
\multicolumn{1}{c}{$\alpha^{T}$\footnotemark[5]}  \\
&&&&&\multicolumn{1}{c}{[b]} &
&\multicolumn{5}{c}{$10^{-10} {\rm Hz}/({\rm V/m})^2$}  \\
\hline
13 & Al  &  27 & 5/2 & 3.6415 & 0.14  & 3 &  0.0158 &  0.2683 & -0.0119 &  0.2563 & \\
   &     &     &     &        &       & 2 & -0.0222 & -0.1073 & -0.0096 & -0.1169 & \\
                     
37 & Rb  &  85 & 5/2 & 1.3530 & 0.27  & 3 &  0.4599 & -0.0073 & -0.0070 & -0.0143 & \\
   &     &     &     &        &       & 2 & -0.6439 &  0.0029 & -0.0056 & -0.0027 & \\
                     
37 & Rb  &  87 & 3/2 & 2.7510 & 0.132 & 2 &  0.9332 & -0.0148 & -0.0034 & -0.0182 & -0.0234 \\
   &     &     &     &        &       & 1 & -1.5554 &  0.0025 & -0.0017 &  0.0007 & 0.0009 \\
                     
55 & Cs  & 133 & 7/2 & 2.5820 &-0.004 & 4 &  1.9770 & -0.0262 & 0.0002 & -0.0260 & -0.034 \\
   &     &     &     &        &       & 3 & -2.5419 &  0.0141 & 0.0002 &  0.0143 & 0.0184 \\
                     
70 & Yb$^+$ & 171 & 1/2 & 0.4940 & 1     & 1 &  0.0844 & -0.0023 &  0 & -0.0023 & \\
   &     &     &     &        &       & 0 & -0.2533 &  0      &  0 &   0 & \\
                     
70 & Yb$^+$ & 173 & 5/2 & -0.6775 & 2.8  & 3 & -0.1158 &  0.0031 & -0.0390 & -0.0359 & \\
   &     &     &     &         &      & 2 &  0.1621 & -0.0012 & -0.0312 & -0.0325 & \\
                     
80 & Hg$^+$ & 199 & 1/2 & -0.5603 & 0.4  & 1 &  0.0271 & -0.0018 &  0  & -0.0018 & \\
   &     &     &     &         &      & 0 & -0.0814 &  0 &  0 &  0 & \\
                     
80 & Hg$^+$ & 201 & 3/2 &  0.5060 & 0.8  & 2 & -0.0300 &  0.0020 & -0.0014 &  0.0005 & \\
   &     &     &     &         &      & 1 &  0.0501 & -0.0003 & -0.0007 & -0.0011 & \\
                     
87 & Fr  & 211 & 9/2 &  4.0005 &-0.19 & 5 &  6.9771 & -0.1459 & 0.0188 & -0.1271 & -0.1633 \\
   &     &     &     &         &      & 4 & -8.5275 &  0.0908 & 0.0175 &  0.1083 & 0.1392 \\
                     
87 & Fr  & 221 & 5/2 & 1.5800 &-0.98 & 3 &  2.7556 & -0.0576 & 0.0970 & 0.0394 & 0.0506 \\
   &     &     &     &        &      & 2 & -3.8579 &  0.0230 & 0.0776 & 0.1006 & 0.129 \\
                     
87 & Fr  & 223 & 3/2 & 1.1700 & 1.17 & 2 &  2.0406 & -0.0427 & -0.1158 & -0.158 & -0.204 \\
   &     &     &     &        &      & 1 & -3.4010 &  0.0071 & -0.0579 & -0.0508 & -0.0653 \\
\end{tabular}
\footnotetext[1]{Reference~\cite{Sto05}}
\footnotetext[2]{Magnetic dipole HFI contribution}
\footnotetext[3]{Electric quadrupole HFI contribution}
\footnotetext[4]{Total tensor polarizability,
  $\alpha^{T}=\alpha_A^{T}+\alpha_B^{T}$}
\footnotetext[5]{Corrected tensor polarizability (see text for discussion)}
\end{ruledtabular}
\end{table*}

Table \ref{t:1} shows the results of calculation of scalar and tensor
hyperfine mediated polarizabilities for two hyperfine components of the ground state
of some isotopes of Al, Rb, Cs, Yb$^+$, Hg$^+$, and Fr. Magnetic
dipole and electric quadrupole HFI contributions to the tensor
polarizabilities ($\alpha^{T}$) are shown separately. Notice that
there is no electric quadrupole contribution to the scalar
polarizabilities ($\alpha^{S}$), as discussed previously.
Another interesting thing to note
is that electric quadrupole does not contribute to the frequency shift
of the hyperfine transition. Although the tensor polarizabilities are
different for states with $F=I+1/2$ and $F=I-1/2$, the energy shifts,
which includes $F$-dependent factors (see Eq.~(\ref{Eq:alphaBreakDown})) are the same. This
is only true for states with $M_F=0$, where $M_F$ is projection of $F$.

The accuracy of the calculations is different for scalar and tensor
polarizabilities, being a few per cent for scalar polarizabilities and
about 30\% for tensor polarizabilities. This is because we include
only Brueckner-type correlations, or correlations which can be reduced
to redefinition of single-electron orbitals. Such approximation works
very well for $s_{1/2}$ and $p_{1/2}$ states which dominate in the
scalar polarizabilities. For tensor polarizabilities large
contribution comes from $p_{3/2}$ and $d_{3/2}$ states where Brueckner
approximation is not so good, especially for the HFI matrix
elements. To get accurate results for tensor polarizabilities one has
to include structure radiation and other non-Brueckner higher-order
correlation corrections. This goes beyond the scope of present work.

To illustrate the accuracy of the calculations we would like to
compare our results to previous calculations and available
experimental data. There were numerous calculations and measurements of
the Stark frequency shift for the hyperfine transitions of the ground state
of Cs, Rb, and other atoms and ions (see, e.g., Ref.~\cite{AngDzuFla06}
and references therein). The results are usually presented in terms of
the coefficient $k$ related to frequency shift by
\begin{equation}
  \delta \Delta E = k\mathcal{E}_0^2 \ ,
\label{eq:k}
\end{equation}
where $\Delta E$ is the energy interval between two hyperfine components of the 
ground state and $\delta \Delta E$ is its change due to electric field 
$\mathcal{E}_0$. One can find the values of $k$ for all atoms and ions
from Table~\ref{t:1} using the polarizabilities from this table and 
expressions (\ref{eq:alpha-E}) and (\ref{Eq:alphaBreakDown}). 
Corresponding values for $^{87}$Rb, $^{133}$Cs, $^{171}$Yb$^+$, and $^{199}$Hg$^+$
are presented in Table~\ref{t:2} and compared to previous calculations of
Ref.~\cite{AngDzuFla06}. Since both calculations are performed with the
same method, we assume the same uncertainty. Some differences in central 
values for Yb and Hg are due to differences in the details of the calculations.
These differences are within the declared uncertainty. 

\begin{table}
\caption{Stark frequency shift: comparison with previous
  calculations.}
\label{t:2}
\begin{ruledtabular}
\begin{tabular}{ccc}
\multicolumn{1}{c}{Atom} &
\multicolumn{2}{c}{$k, \ [10^{-10} {\rm Hz}/({\rm V/m})^2]$} \\
 & This work & Ref.~\cite{AngDzuFla06} \\
\hline
$^{87}$Rb  & -1.24(1)  & -1.24(1)  \\
$^{133}$Cs & -2.26(2)  & -2.26(2)  \\
$^{171}$Yb & -0.167(9) & -0.171(9) \\
$^{199}$Hg & -0.056(3) & -0.060(3) \\
\end{tabular}
\end{ruledtabular}
\end{table}

As has been discussed above the accuracy for the tensor polarizabilities
is lower. Our value of the tensor polarizability for the $F=4$ state of 
cesium is $-2.60$~Hz(V/cm)$^2$. This is about 30\% smaller that the experimental
value $-3.34(2)(25)$~Hz(V/cm)$^2$~\cite{OspRasWeis03} and the semiempirical
value $-3.72(25)$~Hz(V/cm)$^2$~\cite{HofMorUlz08}. The most likely reason for 
the difference is the contribution from the non-Brueckner correlations 
which are not included in present work. Therefore, we assume 30\% uncertainty
for the calculated tensor polarizabilities in present work.

To improve the predicted values for the tensor polarizabilities of Rb and Fr,
which both have electron structure similar to those of cesium, we multiply
the {\em ab initio} results for these atoms by the factor 1.28, which is the 
ratio of the experimental tensor polarizability for Cs to the theoretical one. 
The resulting values are presented in the last column of Table~\ref{t:1}.